\documentclass[lettersize,journal]{IEEEtran}

\usepackage{amsmath,amsfonts}
\usepackage[english]{babel}
\usepackage{siunitx}
\usepackage{mathtools}
\usepackage{orcidlink}

\usepackage{hyperref}
\usepackage[capitalize]{cleveref}
\crefname{enumi}{Assumption}{Assumption}

\usepackage{graphicx}

\usepackage{booktabs}

\usepackage{placeins}
\usepackage{setspace}
\usepackage{cite}
\usepackage{textcomp}

\newcommand\soutpars[1]{\let\helpcmd\sout\parhelp#1\par\relax\relax}
\long\def\parhelp#1\par#2\relax{%
  \helpcmd{#1}\ifx\relax#2\else\par\parhelp#2\relax\fi%
}

\begin{document}

\title{Cryogenic Characterization of Low-Frequency Noise in 40-nm CMOS} % include "and modeling"

\author{\IEEEauthorblockN{
Gerd Kiene\orcidlink{0000-0002-2924-2033}, \IEEEmembership{Member,~IEEE},
Sad{\i}k \.{I}lik \orcidlink{0000-0002-5647-7188},
Luigi Mastrodomenico \orcidlink{0009-0002-5897-7423},
Masoud Babaie\orcidlink{0000-0001-7635-5324}, \IEEEmembership{Senior Member,~IEEE}  and
Fabio Sebastiano\orcidlink{0000-0002-8489-9409}, \IEEEmembership{Senior Member,~IEEE}
}

\thanks{This work was supported by Intel Corporation and NXP Semiconductors.
(Corresponding author: Gerd Kiene.)}%
\thanks{Gerd Kiene, Luigi Mastrodomenico, Sadik Ilik and Fabio Sebastiano are with the Department of Quantum
and Computer Engineering, Delft University of Technology, 2628 CJ Delft,
The Netherlands, and also with Qutech, 2628 CJ Delft, The Netherlands
(e-mail: g.kiene@tudelft.nl).}% <-this % stops a space
\thanks{Masoud Babaie is with the Department of Microelectronics, Delft University
of Technology, 2628 CD Delft, The Netherlands, and also with Qutech,
2628 CJ Delft, The Netherlands.}
}

% make the title area

\markboth{}%
{Shell \MakeLowercase{\textit{et al.}}: A Sample Article Using IEEEtran.cls for IEEE Journals}

%\IEEEpubid{0000--0000/00\$00.00~\copyright~2022 IEEE}

\maketitle

\begin{abstract}
This paper presents an extensive characterization of the low-frequency noise (LFN) at room temperature (RT) and cryogenic temperature (4.2\,K) of 40-nm bulk-CMOS transistors.
The noise is measured over a wide range of bias conditions and geometries to generate a comprehensive overview of LFN in this technology.
While the RT results are in-line with the literature and the foundry models, the cryogenic behavior diverges in many aspects.
These deviations include changes with respect to RT in magnitude and bias dependence that are conditional on transistor type and geometry, and even an additional systematic Lorentzian feature that is common among individual devices.
Furthermore, we find the scaling of the average LFN with the area and its variability to be similar between RT and 4.2\,K, with the cryogenic scaling reported systematically for the first time.
The findings suggest that, as no consistent decrease of LFN at lower temperatures is observed while the white noise is reduced, the impact of LFN for precision analog design at cryogenic temperatures gains a more predominant role.
\end{abstract}


\section{Introduction}

Cryogenic electronics is a rapidly expanding field driven by applications such as quantum computing and space exploration \cite{vandersypen2017interfacing,patterson2006assessment}.
Significant challenges arise in designing CMOS electronics for cryogenic temperatures due to the incomplete understanding of cryogenic transistor behavior, thus causing over-design or even design failures.
To address the situation, extensive characterization and modeling of DC and RF behavior have been performed, e.g., in \cite{beckers2020physical,chakraborty2021characterization}.

Within the cryogenic behavior, noise is generally less explored, but works on both broad-band noise and low-frequency noise have been reported.
In broad-band noise characterization, the noise is scaling less than expected from a purely thermal origin, thus suggesting that shot noise dominates at cryogenic temperatures \cite{ohmori2023variable}.
For designing systems at cryogenic temperature, LFN also plays a crucial role, e.g. for the phase noise of phase-locked loops (PLL) \cite{gong2022cryo} and the input-referred noise of transimpedance amplifiers (TIA) \cite{le2020low}.
Early work on cryogenic LFN provided evidence pointing to both carrier number and mobility fluctuations as possible origins of the noise at low temperatures \cite{hafez1989flicker,aoki1977low}.
In \cite{paz2020performance}, measurements of a modern SOI process over a range of bias points are shown to be compatible with the model including carrier number fluctuations and  correlated mobility fluctuations \cite{ghibaudo1991improved}.
In \cite{asanovski2023understanding}, the lack of the expected decrease of LFN at cryogenic temperatures is attributed to band-tail states acting as traps.
This is corroborated by the same band-tail states possibly causing the saturation of the MOSFET sub-threshold slope at cryogenic temperatures \cite{hafez1990assessment,bohuslavskyi2019cryogenic}.
In \cite{oka2020toward}, characterization and modeling of large-area devices manufactured on different crystal orientations suggest carrier number fluctuation caused by interface traps as an alternative explanation for cryogenic LFN.
While these characterizations, typically only of one or few devices, have added significantly to the understanding of LFN at cryogenic temperature, they can only partially inform design decisions.
%This one is confused and quite weak: \cite{nafaa2018low}

Missing in current literature is a more extensive characterization that can both guide the circuit designers with accurate predictions and help understand the physical origin of LFN.
To fill this gap, we measure several different geometries and characterize them over a wide range of bias voltages, as commonly required for design space explorations.
Multiple individual devices with the same geometry are measured at the same bias, since statistical characterization is crucial for accurate analysis due to the large variations between individual devices, even for large-area devices \cite{da2016physics}, which are already present at RT and hence expected also at cryogenic temperatures.
Such an extensive characterization enables us to also perform a systematic analysis of the scaling of LFN and its variability with device area, for the first time at cryogenic temperature. Since noise exhibits a significant spread over different devices, it has been crucial to extend our study to a significant number of samples over varying geometries to obtain a comprehensive and complete analysis.
% old version
%Furthermore, the analysis of multiple equal devices enabled the discrimination of a Lorentzian feature that is unexpectedly systematic among different devices.
%This feature is dependent on device geometry, bias, and temperature, and causes increased device noise that is detrimental to cryogenic circuits.

%new version
Furthermore, the analysis of multiple equal devices enabled the discrimination of a Lorentzian feature that is unexpectedly systematic among different devices. Previous research, as seen in \cite{ohmori2023variable} and \cite{inaba2023determining}, has noted a dominant Lorentzian signature at cryogenic temperatures. However, due to the lack of a detailed and focused investigation into this signature, it has not been reported as a systematic result. For the first time, we demonstrate that this feature is dependent on device geometry, bias, and temperature, leading to increased device noise that proves detrimental to cryogenic circuits. Thus, these extensive new data both improve the understanding of low-frequency noise and enable circuit designers to get better noise estimations.

\IEEEpubidadjcol

This article is organized as follows.
We begin by introducing the experimental methods in \cref{section:methods}, then present the measurement results and analyze them in \cref{section:characterization}, discuss the results in the light of their impact on precision analog design in \cref{section:discussion} and draw the conclusions in \cref{section:conclusion}.

\section{Methods}
\label{section:methods}
%original text
%The measurement setup was designed to facilitate the characterization of many devices at cryogenic temperature, as necessary to generate the statistics for LFN.
%For this, a chip was fabricated in \SI{40}{\nano\meter} LP CMOS bulk technology (\cref{fig_measurement_setup}\,a), which carried 8 individual transistors for each of the several geometries, for a total of 400 heavily multiplexed devices under test (DUT) per chip.
%In the chip, the devices' source and drain (V\textsubscript{d}, V\textsubscript{s}) are connected in parallel.
%Multiplexing between them is implemented, similar to \cite{t2019subthreshold}, by connecting a single device gate at a time to the gate bias (V\textsubscript{g}), see \cref{fig_measurement_setup}\,b) and c) for the NMOS and PMOS schematic.
%For reliable rail-to-rail operation, multiplexing of V\textsubscript{g} is implemented with thick-oxide transistors, supplied by \SI{2.5}{\volt}.
%All deselected transistors are biased with V\textsubscript{gs}=0, and the source is kept at ground/V\textsubscript{dd} for all NMOS/PMOS measurements.
%Unless otherwise noted, all measured transistors are of low-threshold-voltage (LVT) flavor.

\begin{figure}[t!]
\centering
\includegraphics[width=\linewidth]{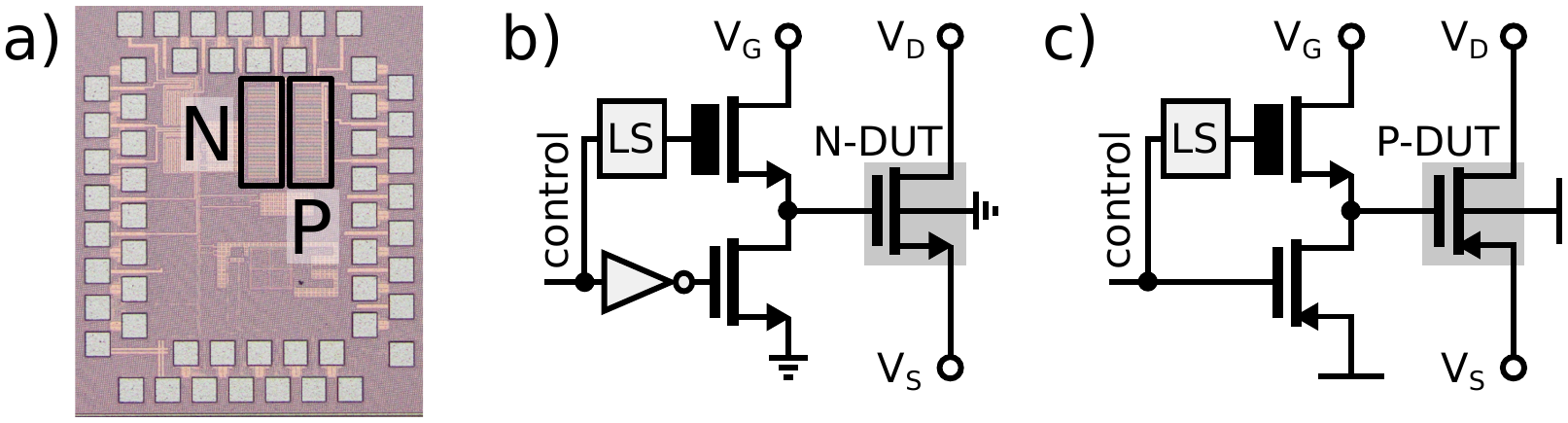}
\caption{a) Micrograph of the test chip;  b) Schematic of each NMOS; c) Schematic of each PMOS. The thick-oxide selection transistors are shown with a thicker-gate symbol.}
\label{fig_measurement_setup}
\end{figure}

The measurement setup was designed to facilitate the characterization of many devices at cryogenic temperature, as necessary to generate the statistics for LFN.
For this, a chip was fabricated in \SI{40}{\nano\meter} LP CMOS bulk technology which carried 8 individual transistors for each of the several geometries, for a total of 400 heavily multiplexed devices under test (DUT) per chip.
In the chip, the devices' source and drain (V\textsubscript{d}, V\textsubscript{s}) are connected in parallel.
Multiplexing between them is implemented, similar to \cite{t2019subthreshold}, by connecting a single device gate at a time to the gate bias (V\textsubscript{g}), see \cref{fig_measurement_setup}\, for the NMOS and PMOS schematic.
For reliable rail-to-rail operation, multiplexing of V\textsubscript{g} is implemented with thick-oxide transistors, supplied by \SI{2.5}{\volt}.
All deselected transistors are biased with V\textsubscript{gs}=0, and the source is kept at ground/V\textsubscript{dd} for all NMOS/PMOS measurements.
Unless otherwise noted, all measured transistors are of low-threshold-voltage (LVT) flavor.

Assembled in a DIP package, the chip is mounted on a PCB placed in a dipstick setup and submerged in liquid helium with a temperature sensor mounted close to the sample.
Such a liquid cooling at \SI{4.2}{\kelvin} results in a stable and accurate definition of the ambient temperature, unlike prior characterization in the vacuum environment of the typical cryogenic probe stations, which may suffer from thermalization issues.
For the RT measurements, the ambient temperature was controlled to within $\pm\SI{2.5}{\kelvin}$ around \SI{295}{\kelvin}.
This setup allows for up to 5 days of continuous measurement at cryogenic temperatures using a single \SI{100}{\liter} helium dewar, without labor-intensive thermal cycles.
The sample was kept mounted in the same dipstick for both RT and cryogenic measurements.
Measurements are performed via a resistive bias-T followed by a transimpedance amplifier (TIA) and digitized by an acquisition card.
The resulting data is accurate over the frequency range  from \SI{1}{\hertz} (or \SI{5}{\hertz}, depending on settings of the bias-T) up to \SI{50}{\kilo\hertz}, limited by the TIA bandwidth.
The reported sweeps over V\textsubscript{gs} were performed in two bias configurations: one in the linear region with fixed V\textsubscript{ds}=\SI{50}{\milli\volt}  and one in an effective ``diode'' connection with V\textsubscript{ds}=V\textsubscript{gs}.
The analyzed data is filtered with a rolling median filter for better visibility.
Further details about the measurement setup and the data processing can be found in the supplementary material.

All data and analysis code used for the figures in this paper is available in \cite{measurement_data}.

%original figure

% \begin{figure}[t!]
% \centering
% \includegraphics[width=\linewidth]{figures/flicker_noise_setup.pdf}
% \caption{Simplified noise measurement setup block diagram. The thick-oxide selection transistors are shown with a thicker-gate symbol.}
% \label{fig_measurement_setup}
% \end{figure}

\section{Characterization results}
\label{section:characterization}

\begin{figure}[t!]
\centering
\includegraphics[width=\linewidth]{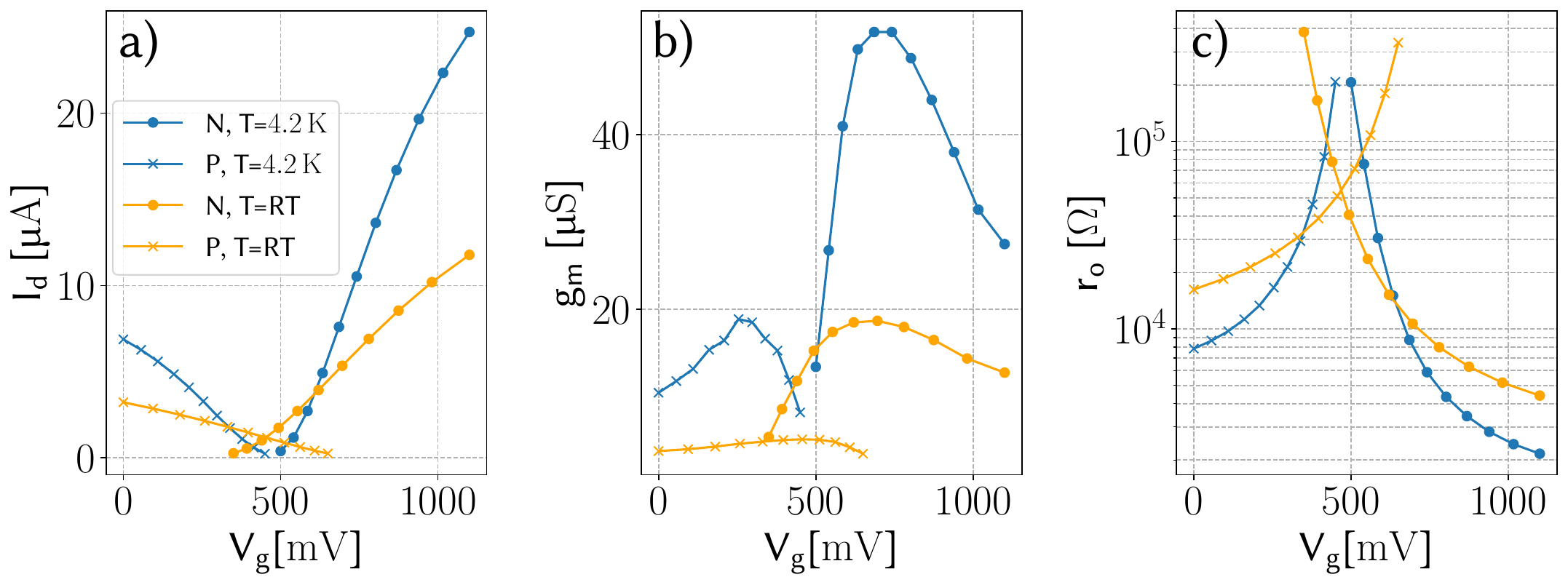}
\caption{Example N/PMOS (\SI{1}{\um}×\SI{1}{\um}) DC-characteristics at V\textsubscript{ds}=\SI{50}{\milli\volt}: a)\, drain current I\textsubscript{d}, b)\,transconductance g\textsubscript{m}, c)\,output resistance r\textsubscript{o}.}
\label{fig_example_device_dc}
\end{figure}

\begin{figure}[t!]
\centering
\includegraphics[width=\linewidth]{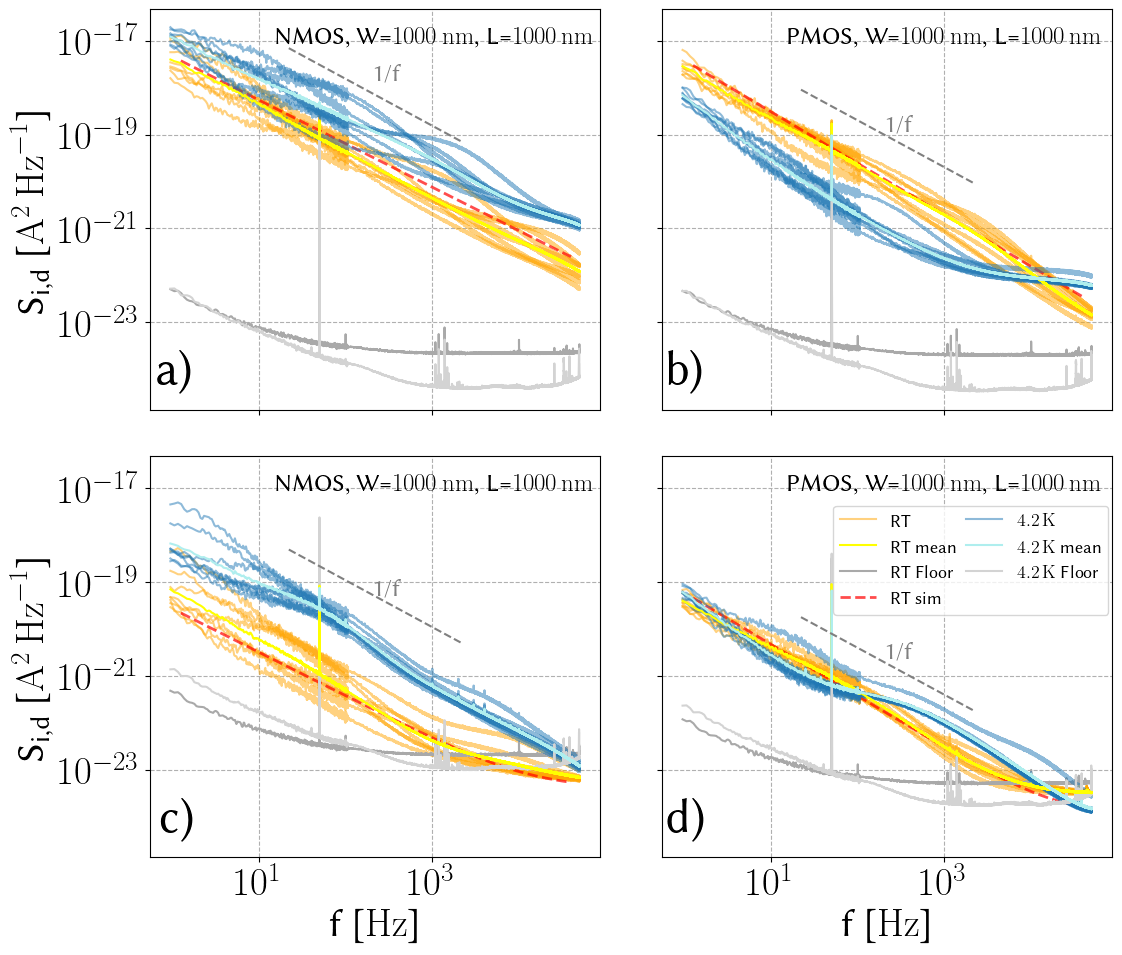}
\caption{Noise spectra for  NMOS (a, c) and  PMOS (b, d) with W$\times$L =\SI{1}{\um}×\SI{1}{\um} at V\textsubscript{gs}=\SI{1.1}{\volt} and V\textsubscript{ds}=\SI{1.1}{\volt} (a,b) and V\textsubscript{ds}=\SI{50}{\milli\volt} (c,d), respectively. 8 devices at the same bias points are shown.}
\label{fig_example_device_sweep}
\end{figure}

\begin{figure*}[t!]
\centering
\includegraphics[width=7.16in]{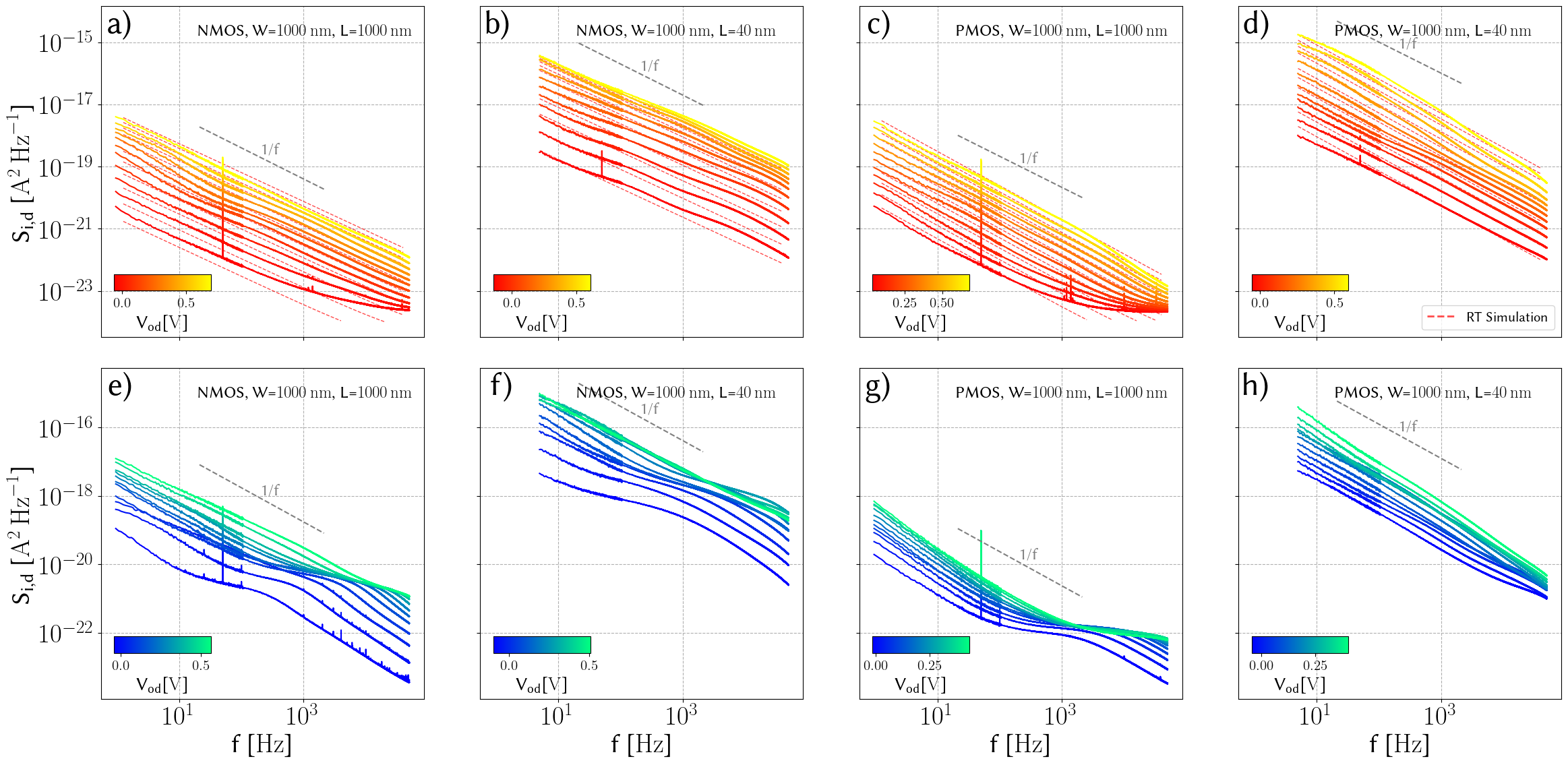}
\caption{Mean noise spectra for PMOS and NMOS devices with W×L=\SI{1}{\micro\meter}×\SI{1}{\micro\meter} and W×L=\SI{1}{\micro\meter}×\SI{40}{\nano\meter}  with V\textsubscript{gs}=V\textsubscript{ds} for logarithmically spaced overdrive-voltage (V\textsubscript{od}=V\textsubscript{gs}-V\textsubscript{th}) at RT (a-d) and  \SI{4.2}{\kelvin} (e-h). Each curve represents the mean of 8 individual devices. The RT plots also show the corresponding simulations of the foundry model with dashed lines.}
\label{fig_bias_sweep_spectra}
\end{figure*}

An example DC characterization of an N/PMOS with W×L=\SI{1}{\micro\meter}×\SI{1}{\micro\meter} is shown in \cref{fig_example_device_dc}.
The characterization shows the typical changes in behavior when moving to cryogenic temperature: the increased threshold voltage (V\textsubscript{th}) and the increased transconductance (g\textsubscript{m}) as well as the reduced output resistance (r\textsubscript{o}) in strong inversion \cite{incandela2018characterization}.

Example noise spectra of eight \SI{1}{\micro\meter}×\SI{1}{\micro\meter} devices at a single bias point V\textsubscript{gs}=V\textsubscript{ds}=\SI{1.1}{\volt} for both N/PMOS are shown in \cref{fig_example_device_sweep}.
The plots also include a conservative noise-floor estimation based on the measured noise floor of the test equipment scaled by the expected additional noise due to the transistor r\textsubscript{o}, further discussed in the supplementary.
Alongside the individual spectra, a logarithmic mean curve is shown, around which individual devices differ significantly at RT, as also reported in \cite{da2016physics}, due to the random distribution of single traps.
This effect is here, for the first time, shown to persist also at cryogenic temperatures.
The RT mean shows typical 1/f behavior and matches the foundry device model well.
The cryogenic results differ, with the NMOS devices showing increased output-referred noise and the PMOS devices showing reduced noise.
While the variability at lower frequencies is qualitatively similar to the RT results, a systematic deviation from the 1/f shape appears in the spectra at higher frequencies.

To substantiate this, we plot in \cref{fig_bias_sweep_spectra} the mean spectra for W×L=\SI{1}{\micro\meter}×\SI{1}{\micro\meter} and W×L=\SI{1}{\micro\meter}×\SI{40}{\nano\meter} NMOS and PMOS devices for 11 logarithmically spaced overdrive-voltage biases (V\textsubscript{od}=V\textsubscript{gs}-V\textsubscript{th}), which are color-coded in the plot.
We concentrate on the mean and do not show the individual devices for better visibility of the effect.
The device geometries were chosen to provide examples of typical devices preferred in analog (high intrinsic gain, long channels) and RF (high speed, short channels) circuits.
For the RT results, we see again that the mean noise approximates a 1/f slope and matches well the foundry device model over a wide range of bias conditions.
The 1/f slope is also visible in the cryogenic results, but a systematic Lorentzian feature appears in the mean curves at higher frequencies.
Signatures of this feature are visible in all the devices (NMOS and PMOS, long and short channels) with the effect appearing at lower frequency in longer devices.
As this effect appears in all the individual devices, it does not average out to a 1/f behavior as otherwise common for random traps. To understand its possible origin, we analyze this Lorentzian in more detail in the following subsection.

\subsection{Systematic Lorentzian}

The systematic Lorentzian feature common to multiple devices and mentioned above has not been reported in prior works. In \cite{ohmori2023variable}, a dominant Lorentzian signature that shows a clear temperature dependence is observed at relatively high drain biases at cryogenic temperatures. In \cite{inaba2023determining}, two traps are identified as slow and fast. They lead to a single Lorentzian signature in low-frequency noise measurement, similar to our results. Still, noise characterization of a single device is presented in these papers, and effects are not investigated extensively and compared with different devices.

To ensure that it is not an artifact of the measurements, we performed additional verifications.
As the equipment and the cabling adopted in the cryogenic setup are identical to the RT setup, for which the Lorentzian does not appear, we specifically investigate the components cooled to cryogenic temperature, including the chip itself and the associated passive components (resistors, capacitors) on the measurement board.
Leakage currents from the deselected transistors are excluded by recording spectra with all transistors deselected and sweeping the drain bias.
The on-chip multiplexing was tested by reducing the thick-oxide switch supply by \SI{500}{\milli\volt} without measurable effects on either the DUT's DC behavior or its noise spectrum. 
The on-board tantalum capacitors have been excluded by measuring the DUT on a board without populating those, resulting in increased interference but the same effect visible.
For testing the thin-film resistors, we replaced the chip with resistors and ran noise acquisitions under various bias conditions, without observing measurable LFN contributed by the resistors.
The additional verifications are discussed in more detail in the supplementary.

\begin{figure}[t!]
\centering
\includegraphics[width=\linewidth]{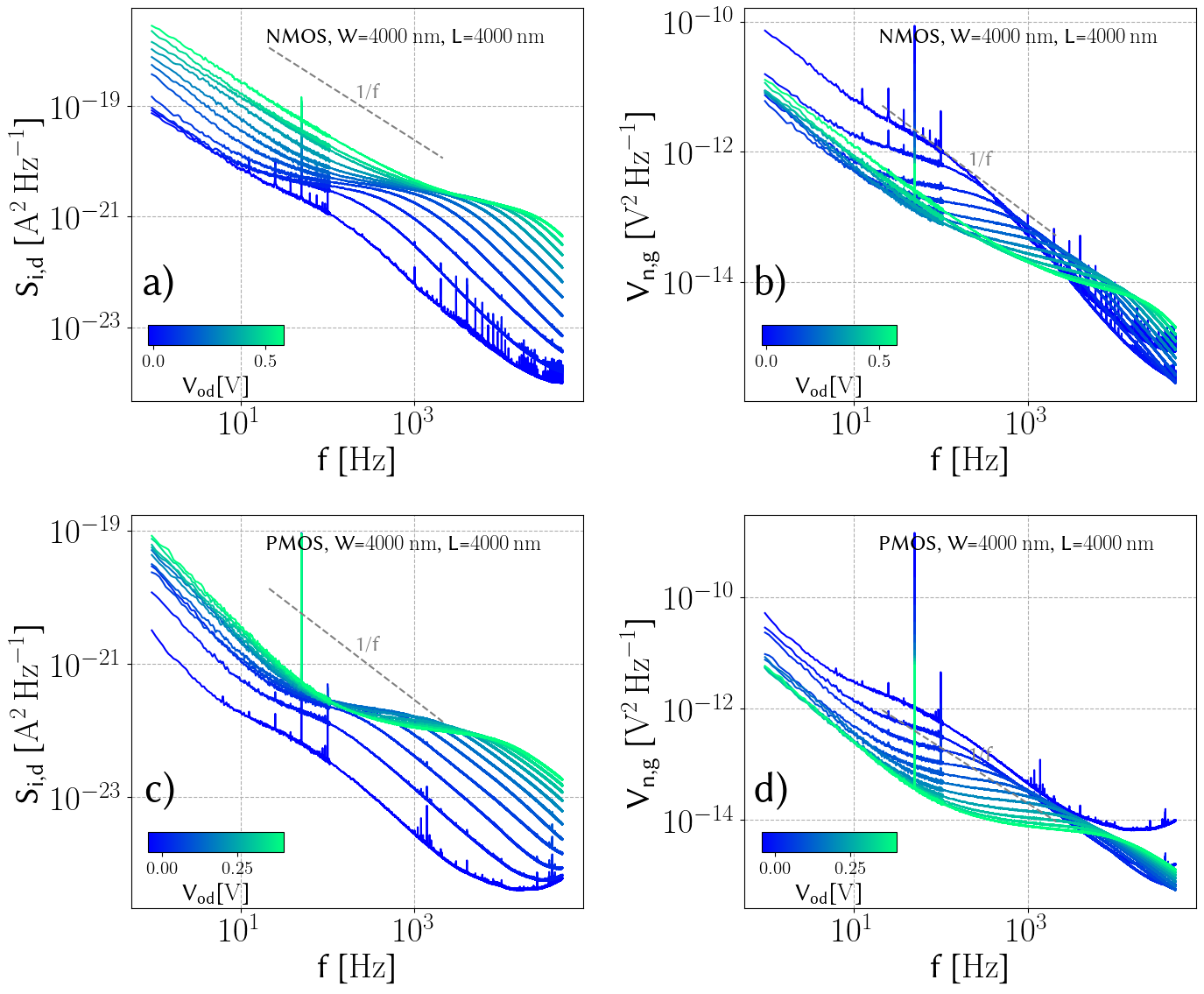}
\caption{Mean noise spectra at \SI{4.2}{\kelvin}. The output (a, c) and input-referred (b, d) of eight N/PMOS devices with WxL=\SI{4}{\micro\meter}x\SI{4}{\micro\meter} are shown, to demonstrate the bias dependence of the systematic Lorentzian.}
\label{fig_lorenzian_plot}
\end{figure}

\begin{figure}[t!]
\centering
\includegraphics[width=\linewidth]{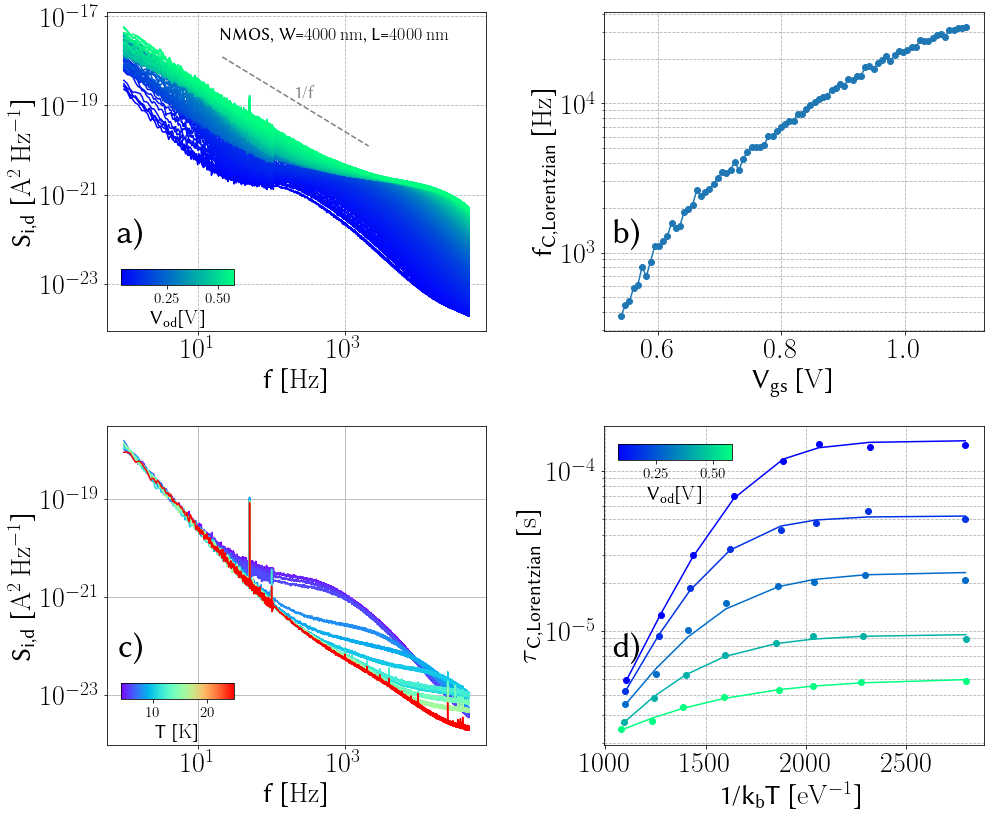}
\caption{Detailed measurements on a W×L=\SI{4}{\micro\meter}×\SI{4}{\micro\meter} NMOS device: a) fine bias sweep at \SI{4.2}{\kelvin}, b) extracted corner frequency of the Lorentzian at \SI{4.2}{\kelvin}, c) temperature sweep for $V_{od}=V_{gs}-V_{th}=\SI{80}{\milli\volt}$, d) extracted time constant of the Lorentzian.}
\label{fig_lorenzian_in_detail}
\end{figure}

To gain a deeper understanding of the Lorentzian effect, we characterized large-area (\SI{4}{\micro\meter}×\SI{4}{\micro\meter}) NMOS and PMOS devices (\cref{fig_lorenzian_plot}), since they show a relatively low mismatch and the Lorentzian feature appears more distinctly compared to small devices, as visible in \cref{fig_bias_sweep_spectra}.
\Cref{fig_lorenzian_plot} shows that the Lorentzian's corner moves to higher frequencies for higher overdrive while V\textsubscript{ds}=V\textsubscript{gs}.
%, and its magnitude in the input-referred spectrum decreases with increasing bias
As we observe negligible individual differences in these large devices concerning the Lorentzian, we performed, in the following, a more detailed series of measurements on a single device, which is shown in \cref{fig_lorenzian_in_detail}.

\Cref{fig_lorenzian_in_detail}\,a) shows a fine sweep of  V\textsubscript{od} of a single device in diode connection, from which the corner frequency $f_{C,Lorentzian}$ as function of the overdrive voltage has been extracted and plotted in \cref{fig_lorenzian_in_detail}\,b).
The strong dependence of the corner frequency with the bias can be attributed to barrier lowering or barrier thinning effects \cite{rogers1985nature,michl2020quantum}.

To further investigate the nature of the effect, the temperature dependence of the trap at a given voltage bias has been measured by varying the position of the DUT mounted in the dipstick with respect to the liquid-helium surface.
It is important to note that, while the temperature sensor was glued as close as possible to the sample on the ceramic package for this sweep, the accuracy in the thermalization of the DUT cannot be precisely quantified for the adopted setup above \SI{4.2}{\kelvin}.
When sweeping the temperature from \SI{4.2}{\kelvin} to \SI{25}{\kelvin}, the Lorentzian moves to higher frequencies in diode connection, see \cref{fig_lorenzian_in_detail}c).
%Above \SI{12}{\kelvin} we could not detect the feature anymore.
Above \SI{12}{\kelvin}, the corner frequency exceeds the upper bandwidth limit of the setup.

Assuming the presence of a trap, the trap time-constant $\tau_{C,Lorentzian}=1/(2\pi f_{C,Lorentzian})$ is extracted and plotted in \cref{fig_lorenzian_in_detail} d) versus the inverse of the thermal energy. The saturation of the time constant at low temperatures suggests the existence of a temperature-independent contribution similar to that observed and explained in \cite{michl2021efficientpartI,michl2021efficientpartII}.
%The saturation of the time constant at low temperatures suggests the existence of a temperature-independent contribution due to elastic tunneling.
As in \cref{fig_lorenzian_in_detail} b), a decrease in the time-constant can be observed with increasing overdrive voltage.
%The theory of inelastic tunneling into traps in \cite{kirton1989noise} suggests a temperature dependence of the trap time-constant: 
%For thermally activated inelastic tunneling, \cite{kirton1989noise} suggests a temperature dependence of the trap time-constant: 
For thermally activated contribution, \cite{kirton1989noise} suggests a temperature dependence of the trap time-constant according to:
%equation 2.27b and also around equation 5.4
\begin{equation}
    \tau_{thermal}=\tau_0 e^{\frac{E_b}{k_bT}}
\end{equation}
where $\tau_0=\frac{1}{\sigma_0vn}$ is depending on the cross-section pre-factor $\sigma_0$, the carrier velocity $v$ and channel carrier concentration $n$, $E_b$ is the energy barrier and $k_b$ the Boltzmann constant.
% original text
%Assuming a temperature independent contribution of elastic tunneling with time constant $\tau_{inel}$ \cite{michl2021evidence}, the combined effect can be described with \cite{rogers1985nature}:
%Assuming a temperature-independent contribution of elastic tunneling, with time constant $\tau_{el}$ \cite{michl2021evidence}, the combined effect can be described as \cite{rogers1985nature}:
Assuming a temperature-independent contribution, with time constant $\tau_{tun}$, the combined effect can be described as in \cite{rogers1985nature,michl2021evidence}:
%\begin{equation}
%  \frac{1}{\tau_{C, Lorentzian}} = \frac{1}{\tau_{thermal}} + %\frac{1}{\tau_{el}}
%    \label{equ_tau}
%\end{equation}
\begin{equation}
    \frac{1}{\tau_{C, Lorentzian}} = \frac{1}{\tau_{thermal}} + \frac{1}{\tau_{tun}}
    \label{equ_tau}
\end{equation}
%
% original text: Depending on bias the trap distance from the Fermi level
%moves from \SI{6}{\milli\eV} to \SI{2}{\milli\eV}, for the initial barrier we estimate a height of \SI{10}{\milli\eV}.
Fitting \cref{equ_tau} to the measured data in \cref{fig_lorenzian_in_detail}\,d) we can determine the corresponding energy barrier $E_b$.
The extracted $E_b$ shifts from \SI{5.9}{\milli\eV} to \SI{2.3}{\milli\eV} for a V\textsubscript{gs} bias from \SI{600}{\milli\volt} to \SI{1100}{\milli\volt}. 
By assuming a linear field dependence of $E_b$ as in \cite{rogers1985nature,michl2020quantum}, $E_b$ can be extrapolated to lie on the order of \SI{10}{\milli\eV} when all terminals are at \SI{0}{\volt}.

%Depending on bias, the $E_b$ shifts between \SI{6}{\milli\eV} to \SI{2}{\milli\eV}, with an initial value estimated in the order of \SI{10}{\milli\eV} at \SI{0}{\volt} bias voltage.

A possible physical cause for this phenomenon could be found in the density of states (DoS) showing an additional distinct Gaussian peak below the conduction band at a doping-dependent activation energy \cite{schenk2006physical}. These extra states in the DoS enhance the probability of state occupation at the respective energy levels, thus elevating the transition rate for that energy level and creating a dominant effect. A doping density in the order of \SI{e18}{\per\cubic\centi\metre} to \SI{e19}{\per\cubic\centi\metre}, which is typical doping for the lightly doped drain regions of a \SI{40}{\nano\meter} bulk CMOS technology, would lead to barrier energy in the order of \SI{10}{\milli\eV} \cite{schenk2006physical}.
This hypothesis is compatible with the observed behavior, including the commonality among different devices, the extracted barrier potential, and the increased magnitude of the effect in saturation compared to the triode (as shown by comparing \cref{fig_example_device_sweep}a,b and \cref{fig_example_device_sweep}c,d).

% Original text
%As the observed effect is systematic over multiple individual devices of the same geometry, randomly distributed oxide traps are    unlikely explanation for the results.
%A possible physical cause for this phenomenon could be found in the band-tail: depending on doping, the density of states (DoS) might not only show an exponential tail \cite{bohuslavskyi2019cryogenic}, but also an additional distinct Gaussian peak below the conduction band at a doping-dependent activation energy \cite{schenk2006physical}.
%Normally, states in this tail would be thermally ionized, but at cryogenic temperatures they might act as traps, possibly explaining the substantiation of noise at these temperatures \cite{asanovski2023understanding}.
%A doping density in the order of \SI{e18}{\per\cubic\centi\metre} to \SI{e19}{\per\cubic\centi\metre}, which is a typical doping for the lightly doped drain regions of a \SI{40}{\nano\meter} bulk CMOS technology, would lead to a barrier energy in the order of \SI{10}{\milli\eV} \cite{schenk2006physical}. 
%This hypothesis is compatible with the observed behavior, including the commonality among different devices, the extracted barrier potential, and the increased magnitude of the effect in saturation compared to triode.

However, to gain a more comprehensive understanding of this phenomenon, further research is required, for instance by testing devices with several different precisely known doping levels and by characterizing other bulk technologies under well-controlled thermal conditions, e.g. by employing liquid helium cooling, and over significant device statistics.

\subsection{Bias dependence}

\begin{figure*}[t!]
\centering
\includegraphics[width=7.16in]{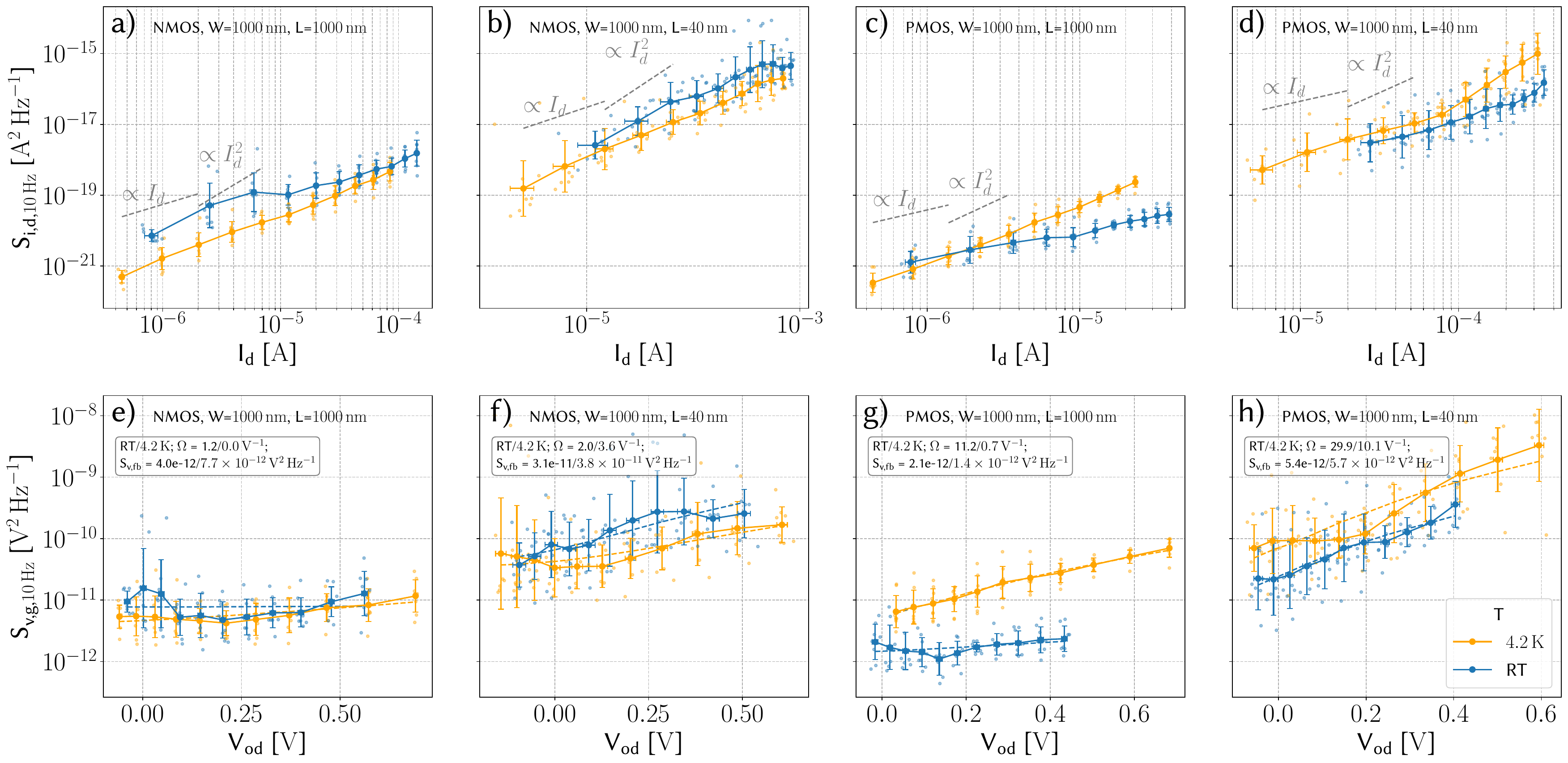}
\caption{Output-referred (a-d) and  input-referred (e-h) noise power spectral density at \SI{10}{\hertz} for N/PMOS at V\textsubscript{gs}=V\textsubscript{ds} over different bias. Each point represents the measurement from an individual device, while solid lines indicate the mean over the device for each voltage bias.}
\label{fig_bias_sweep}
\end{figure*}

At low frequencies, the noise behavior is largely unaffected by the systematic Lorenzian feature and lends itself to a more traditional analysis.
To this end, we sample the data in \cref{fig_bias_sweep_spectra} at \SI{10}{\hertz} and plot the  current noise versus the bias drain current (\cref{fig_bias_sweep} a-d)) and the input-referred noise over V\textsubscript{od} ((\cref{fig_bias_sweep}  e-h) in \cref{fig_bias_sweep}. 
The flicker noise is the dominant noise type in this frequency both at room temperature and \SI{4.2}{\kelvin}.
As expected, we observe a significant variation in the noise of the individual devices around the mean and a drop in the input-referred noise voltage with larger device size, which are analyzed in details in the next subsection.
The output current noise scales with current, with proportionalities ranging from $I_d$ to $I_d^2$.
Generally, the NMOS results are found to scale roughly $\propto I_d$, while the PMOS results are better described with $\propto I_d^2$.
The input-referred data in \cref{fig_bias_sweep} can be qualitatively well-described by
\begin{equation}
    S_{v,g} = \left(1+\Omega\frac{I_d}{g_m}\right)^2 S_{fb}
\label{general_noise_eq}
\end{equation}
with $S_{fb}$ the flat-band voltage spectral density and $\Omega$ the coefficient of the correlated mobility fluctuation  \cite{ghibaudo1991improved}.

Using the Boltzmann statistics, the temperature dependence of \cref{general_noise_eq}, scales linearly with T.
This qualitative correspondence is in-line with the observations in \cite{paz2020performance}.
The extent of the correlated mobility fluctuations changes with device type and geometry.
For the large-area NMOS in e) there are little correlated mobility fluctuations both at RT and cryogenic temperatures, while for the small-area PMOS in h) the effect of correlated mobility is significant at both temperatures.
Furthermore, it is interesting to note the significant differences between the RT and cryogenic results for the large-area PMOS in g): the RT results show a strong correlated mobility contribution, which decreases significantly in the cryogenic results.
The noise magnitude is similar at RT and \SI{4.2}{\kelvin} for plots e), f) and h), with only the large-area PMOS in g) showing a significant drop in input-referred noise at the same overdrive.
The differences in scaling with bias between NMOS and PMOS suggest that the noise is generated via different, possibly doping-dependent, noise mechanisms.
Explaining the generally small difference between the RT and the cryogenic behavior with an increase in interface defect states would demand unlikely high values for the defect density.
This again suggests band-tail states as a possible cause of the substantiated noise as discussed in \cite{asanovski2023understanding}.
A possible further explanation can be found in \cite{ghibaudo2022fluctuation}, with the conclusion that Boltzmann statistics can not be applied, and therefore an effective temperature needs to be introduced, that partially explains the reduced scaling with temperature.

In subthreshold (for negative V\textsubscript{od}), we would expect a constant characteristic in e)-h), assuming the flat-band voltage spectral density is constant.
Although we only probed the edge of this regime, we can observe a tendency for increased input-referred noise in the triode measurements, see the supplementary.
This diverging input-referred voltage in subthreshold can be attributed to mobility fluctuations \cite{ghibaudo1991improved}, which could be confirmed by further characterization in the deep-subthreshold regime.

\subsection{Area dependence}

\begin{figure}[t!]
\centering
\includegraphics[width=3.5in]{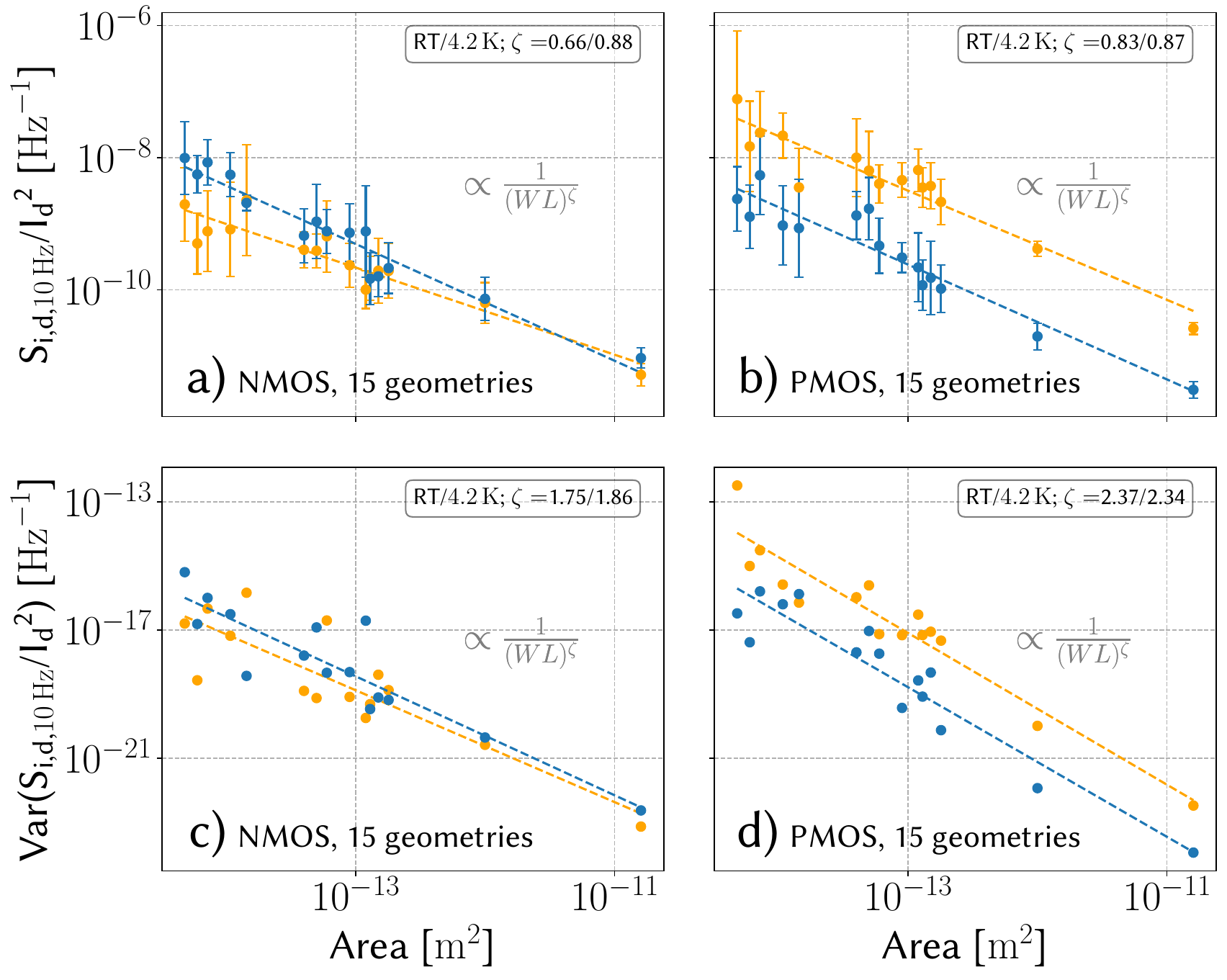}
\caption{a,b) Normalized current noise versus device area for (a) NMOS and (b) PMOS at V\textsubscript{ds}=V\textsubscript{gs}=\SI{1.1}{\volt} for 15 different geometries. c,d) Variance of normalized drain current for NMOS (c) and PMOS (d) at same bias point as in a,b.}
\label{fig_area_sweep}
\end{figure}

\begin{table}[!t]
\caption{Device sizes used in area sweep}
\label{fig_core_sizing}
\centering
   % \ra{1.3}
   \begin{tabular}{cc|cc}
       \toprule
       W [\si{\nano\meter}] & L [\si{\nano\meter}] & W [\si{\nano\meter}] & L [\si{\nano\meter}]  \\
       \midrule
       120 & 40 & 1000 & 60  \\
       120 & 50 & 1000 & 90  \\
       120 & 60 & 1000 & 120  \\
       120 & 90 & 1000 & 150  \\
       120 & 120 & 1000 & 180  \\
       360 & 360 & 1000 & 1000  \\
       1000 & 40 & 4000 & 4000  \\
       1000 & 50 \\
       \bottomrule
   \end{tabular}
\end{table}

At RT, it was shown in \cite{da2016physics} that significant variability is present even in larger devices for the cases of a non-uniform channel.
At cryogenic temperature, this area dependence of LFN noise has not been systematically explored before.
Here we report measurements over area from 8 devices per each of 15 different geometries for both PMOS and NMOS, showing both the normalized current noise (\cref{fig_area_sweep} a)/b) and its variance (\cref{fig_area_sweep} c)/d) as a function of device area.
We observe that the noise scaling of NMOS devices shows generally little behavior differences at RT and cryogenic temperature in both expected value and the variance.
While we do observe a less than proportional scaling with area for the normalized noise at RT in \cref{fig_area_sweep} a), at \SI{4.2}{\kelvin} that scaling is closer to the expected inverse proportionality. 
However, this difference is not highly significant in light of the device variability.
In contrast to this, the PMOS devices in \cref{fig_area_sweep}\,b) show significant reduction in normalized noise and variance compared to RT.
For PMOS, scaling of the noise is very similar over temperature and close to $1/WL$.
For both NMOS and PMOS, the variance scaling in saturation is comparatively weak compared to the cubic dependence expected in the linear region, as explained in \cite{da2016physics} with effects caused by the halo-implants.
The variance scaling exponent with area is larger for PMOS than for NMOS.
At RT, the measurements in the linear region, shown in the supplementary, are closer to the expected scaling of mean and variance with $1/(WL)$ and $1/(WL)^3$ \cite{da2016physics}, but the cryogenic results differ significantly due to influence of the systematic Lorentzian.
As shown in the supplementary, different layout effects were also investigated, but neither  metallization on top of the DUTs nor the presence of dummy transistors affected the results shown here.

\section{Discussion}
\label{section:discussion}

The observations made here demonstrate a wide variety of differences for transistor LFN between RT and cryogenic temperatures.
The differences in behavior are strongly dependent on transistor type, geometry, and bias.

For analog designers adopting transistors in saturation, often in moderate inversion, the changes are however minute, given the radical change in temperature, with the input-referred noise being almost unaffected by temperature for most geometries.
The expected scaling with temperature predicted in \cite{ghibaudo1991improved} is not observed.
A minor increase in input-referred noise can be observed for the NMOS devices, while the minimum-length PMOS devices show a slight decrease.
Only the long PMOS devices significantly benefit from cooling to cryogenic temperatures.
However,  the significant  advantage in lower LFN for longer PMOS devices needs to be traded against the difficulties associated with a larger increase in threshold voltage for PMOS devices, compared to their NMOS counterparts \cite{incandela2018characterization}, also visible in \cref{fig_example_device_dc}. As LFN is to first order invariant to temperature in most device configurations, the  decrease of the white noise \cite{ohmori2023variable} results in an expected large increase in frequency of the flicker noise corner.
This may require the adoption of techniques for LFN mitigation \cite{enz1996circuit} for a wider range of frequencies than at RT.

The systematic Lorentzian further complicates the picture at cryogenic temperatures.
Over a significant frequency range, the additional noise from the Lorentzian dominates the transistors noise behavior.
This in turn requires changes the analysis of input-referred noise in amplifiers and the phase noise in oscillators to reflect the device behavior.
Furthermore, if using techniques like auto-zeroing or chopping, the distinct spectral shape of the Lorentzian noise requires additional rejection at higher frequency than would otherwise be necessary.

\section{Conclusion}
\label{section:conclusion}

The extensive study of LFN in MOS transistors presented in this paper, comprising a wide range of bias points and device geometries at both RT and \SI{4.2}{\kelvin}, uncovers a diverse set of effects and dependencies.
Most prominently, we observed and described a systematic Lorentzian spectrum appearing at cryogenic temperatures.
Apart from this, we observed remarkably constant behavior with temperature for the bias-dependence of the NMOS, and reductions in input-referred noise for the PMOS.
Furthermore, the area scaling was shown to be intact at cryogenic temperature.
Since the observed changes have a significant impact on analog design at cryogenic temperatures, the reported characterization will aid the designers of the next-generation cryogenic systems.

% use section* for acknowledgment
\section*{Acknowledgment}
The authors would like to thank Intel Corp. and NXP Semiconductors for funding, Europractice for MPW services, Atef Akhnoukh for technical support and Ramon Overwater for helpful discussions.

\bibliographystyle{IEEEtran}
\bibliography{gerds_bib}

% Generated by IEEEtran.bst, version: 1.14 (2015/08/26)
\begin{thebibliography}{10}
\providecommand{\url}[1]{#1}
\csname url@samestyle\endcsname
\providecommand{\newblock}{\relax}
\providecommand{\bibinfo}[2]{#2}
\providecommand{\BIBentrySTDinterwordspacing}{\spaceskip=0pt\relax}
\providecommand{\BIBentryALTinterwordstretchfactor}{4}
\providecommand{\BIBentryALTinterwordspacing}{\spaceskip=\fontdimen2\font plus
\BIBentryALTinterwordstretchfactor\fontdimen3\font minus \fontdimen4\font\relax}
\providecommand{\BIBforeignlanguage}[2]{{%
\expandafter\ifx\csname l@#1\endcsname\relax
\typeout{** WARNING: IEEEtran.bst: No hyphenation pattern has been}%
\typeout{** loaded for the language `#1'. Using the pattern for}%
\typeout{** the default language instead.}%
\else
\language=\csname l@#1\endcsname
\fi
#2}}
\providecommand{\BIBdecl}{\relax}
\BIBdecl

\bibitem{vandersypen2017interfacing}
L.~Vandersypen, H.~Bluhm, J.~Clarke, A.~Dzurak, R.~Ishihara, A.~Morello, D.~Reilly, L.~Schreiber, and M.~Veldhorst, ``Interfacing spin qubits in quantum dots and donors—hot, dense, and coherent,'' \emph{npj Quantum Information}, vol.~3, no.~1, p.~34, 2017.

\bibitem{patterson2006assessment}
R.~Patterson, A.~Hammoud, and M.~Elbuluk, ``Assessment of electronics for cryogenic space exploration missions,'' \emph{Cryogenics}, vol.~46, no. 2-3, pp. 231--236, 2006.

\bibitem{beckers2020physical}
A.~Beckers, F.~Jazaeri, A.~Grill, S.~Narasimhamoorthy, B.~Parvais, and C.~Enz, ``Physical model of low-temperature to cryogenic threshold voltage in mosfets,'' \emph{IEEE Journal of the Electron Devices Society}, vol.~8, pp. 780--788, 2020.

\bibitem{chakraborty2021characterization}
W.~Chakraborty, K.~A. Aabrar, J.~Gomez, R.~Saligram, A.~Raychowdhury, P.~Fay, and S.~Datta, ``Characterization and modeling of 22 nm fdsoi cryogenic rf cmos,'' \emph{IEEE Journal on Exploratory Solid-State Computational Devices and Circuits}, vol.~7, no.~2, pp. 184--192, 2021.

\bibitem{ohmori2023variable}
K.~Ohmori and S.~Amakawa, ``Variable-temperature broadband noise characterization of mosfets for cryogenic electronics: From room temperature down to 3 k,'' in \emph{2023 7th IEEE Electron Devices Technology \& Manufacturing Conference (EDTM)}.\hskip 1em plus 0.5em minus 0.4em\relax IEEE, 2023, pp. 1--3.

\bibitem{gong2022cryo}
J.~Gong, E.~Charbon, F.~Sebastiano, and M.~Babaie, ``A cryo-cmos pll for quantum computing applications,'' \emph{IEEE Journal of Solid-State Circuits}, 2022.

\bibitem{le2020low}
L.~Le~Guevel, G.~Billiot, B.~Cardoso~Paz, M.~Tagliaferri, S.~De~Franceschi, R.~Maurand, M.~Cass{\'e}, M.~Zurita, M.~Sanquer, M.~Vinet \emph{et~al.}, ``Low-power transimpedance amplifier for cryogenic integration with quantum devices,'' \emph{Applied Physics Reviews}, vol.~7, no.~4, p. 041407, 2020.

\bibitem{hafez1989flicker}
I.~Hafez, G.~Ghibaudo, and F.~Balestra, ``Flicker noise in metal-oxide-semiconductor transistors from liquid helium to room temperature,'' \emph{Journal of applied physics}, vol.~66, no.~5, pp. 2211--2213, 1989.

\bibitem{aoki1977low}
M.~Aoki, H.~Katto, and E.~Yamada, ``Low-frequency 1/f noise in {MOSFET}’s at low current levels,'' \emph{Journal of Applied Physics}, vol.~48, no.~12, pp. 5135--5140, 1977.

\bibitem{paz2020performance}
B.~C. Paz, M.~Cass{\'e}, C.~Theodorou, G.~Ghibaudo, T.~Kammler, L.~Pirro, M.~Vinet, S.~de~Franceschi, T.~Meunier, and F.~Gaillard, ``Performance and low-frequency noise of 22-nm {FDSOI} down to 4.2 {K} for cryogenic applications,'' \emph{IEEE Transactions on Electron Devices}, vol.~67, no.~11, pp. 4563--4567, 2020.

\bibitem{ghibaudo1991improved}
G.~Ghibaudo, O.~Roux, C.~Nguyen-Duc, F.~Balestra, and J.~Brini, ``Improved analysis of low frequency noise in field-effect {MOS} transistors,'' \emph{physica status solidi (a)}, vol. 124, no.~2, pp. 571--581, 1991.

\bibitem{asanovski2023understanding}
R.~Asanovski, A.~Grill, J.~Franco, P.~Palestri, A.~Beckers, B.~Kaczer, and L.~Selmi, ``{U}nderstanding the {E}xcess 1/f {N}oise in {MOSFET}s at {C}ryogenic {T}emperatures,'' \emph{IEEE Transactions on Electron Devices}, 2023.

\bibitem{hafez1990assessment}
I.~Hafez, G.~Ghibaudo, and F.~Balestra, ``Assessment of interface state density in silicon metal-oxide-semiconductor transistors at room, liquid-nitrogen, and liquid-helium temperatures,'' \emph{Journal of applied physics}, vol.~67, no.~4, pp. 1950--1952, 1990.

\bibitem{bohuslavskyi2019cryogenic}
H.~Bohuslavskyi, A.~Jansen, S.~Barraud, V.~Barral, M.~Cass{\'e}, L.~Le~Guevel, X.~Jehl, L.~Hutin, B.~Bertrand, G.~Billiot \emph{et~al.}, ``Cryogenic subthreshold swing saturation in {FD-SOI} {MOSFET}s described with band broadening,'' \emph{IEEE Electron Device Letters}, vol.~40, no.~5, pp. 784--787, 2019.

\bibitem{oka2020toward}
H.~Oka, T.~Matsukawa, K.~Kato, S.~Iizuka, W.~Mizubayashi, K.~Endo, T.~Yasuda, and T.~Mori, ``Toward long-coherence-time {S}i spin {Q}ubit: {T}he origin of low-frequency noise in cryo-{CMOS},'' in \emph{2020 IEEE Symposium on VLSI Technology}.\hskip 1em plus 0.5em minus 0.4em\relax IEEE, 2020, pp. 1--2.

\bibitem{da2016physics}
M.~B. da~Silva, H.~P. Tuinhout, A.~Zegers-van Duijnhoven, G.~I. Wirth, and A.~J. Scholten, ``A physics-based statistical {RTN} model for the low frequency noise in {MOSFET}s,'' \emph{IEEE Transactions on Electron Devices}, vol.~63, no.~9, pp. 3683--3692, 2016.

\bibitem{inaba2023determining}
T.~Inaba, H.~Oka, H.~Asai, H.~Fuketa, S.~Iizuka, K.~Kato, S.~Shitakata, K.~Fukuda, and T.~Mori, ``Determining the low-frequency noise source in cryogenic operation of short-channel bulk mosfets,'' in \emph{2023 IEEE Symposium on VLSI Technology and Circuits (VLSI Technology and Circuits)}.\hskip 1em plus 0.5em minus 0.4em\relax IEEE, 2023, pp. 1--2.

\bibitem{t2019subthreshold}
P.~t~Hart, M.~Babaie, E.~Charbon, A.~Vladimirescu, and F.~Sebastiano, ``Subthreshold mismatch in nanometer cmos at cryogenic temperatures,'' in \emph{ESSDERC 2019-49th European Solid-State Device Research Conference (ESSDERC)}.\hskip 1em plus 0.5em minus 0.4em\relax IEEE, 2019, pp. 98--101.

\bibitem{measurement_data}
G.~Kiene and S.~Ilik, ``Data for: {L}ow frequency noise characterization in 40nm {C}ryo-{CMOS},'' \emph{doi 10.4121/c41c26f7-586f-48ca-8abe-9127b2d97c60}, 2023.

\bibitem{incandela2018characterization}
R.~M. Incandela, L.~Song, H.~Homulle, E.~Charbon, A.~Vladimirescu, and F.~Sebastiano, ``Characterization and compact modeling of nanometer {CMOS} transistors at deep-cryogenic temperatures,'' \emph{IEEE Journal of the Electron Devices Society}, vol.~6, pp. 996--1006, 2018.

\bibitem{rogers1985nature}
C.~Rogers and R.~Buhrman, ``Nature of single-localized-electron states derived from tunneling measurements,'' \emph{Physical review letters}, vol.~55, no.~8, p. 859, 1985.

\bibitem{michl2020quantum}
J.~Michl, A.~Grill, D.~Claes, G.~Rzepa, B.~Kaczer, D.~Linten, I.~Radu, T.~Grasser, and M.~Waltl, ``Quantum mechanical charge trap modeling to explain {BTI} at cryogenic temperatures,'' in \emph{2020 IEEE International Reliability Physics Symposium (IRPS)}.\hskip 1em plus 0.5em minus 0.4em\relax IEEE, 2020, pp. 1--6.

\bibitem{michl2021efficientpartI}
J.~Michl, A.~Grill, D.~Waldhoer, W.~Goes, B.~Kaczer, D.~Linten, B.~Parvais, B.~Govoreanu, I.~Radu, M.~Waltl \emph{et~al.}, ``Efficient modeling of charge trapping at cryogenic temperatures—part i: Theory,'' \emph{IEEE Transactions on Electron Devices}, vol.~68, no.~12, pp. 6365--6371, 2021.

\bibitem{michl2021efficientpartII}
J.~Michl, A.~Grill, D.~Waldhoer, W.~Goes, B.~Kaczer, D.~Linten, B.~Parvais, B.~Govoreanu, I.~Radu, T.~Grasser \emph{et~al.}, ``Efficient modeling of charge trapping at cryogenic temperatures—part ii: Experimental,'' \emph{IEEE Transactions on Electron Devices}, vol.~68, no.~12, pp. 6372--6378, 2021.

\bibitem{kirton1989noise}
M.~Kirton and M.~Uren, ``Noise in solid-state microstructures: A new perspective on individual defects, interface states and low-frequency (1/f) noise,'' \emph{Advances in Physics}, vol.~38, no.~4, pp. 367--468, 1989.

\bibitem{michl2021evidence}
J.~Michl, A.~Grill, B.~Stampfer, D.~Waldhoer, C.~Schleich, T.~Knobloch, E.~Ioannidis, H.~Enichlmair, R.~Minixhofer, B.~Kaczer \emph{et~al.}, ``Evidence of {T}unneling {D}riven {R}andom {T}elegraph {N}oise in {C}ryo-{CMOS},'' in \emph{2021 IEEE International Electron Devices Meeting (IEDM)}.\hskip 1em plus 0.5em minus 0.4em\relax IEEE, 2021, pp. 31--3.

\bibitem{schenk2006physical}
A.~Schenk, P.~P. Altermatt, and B.~Schmithusen, ``Physical model of incomplete ionization for silicon device simulation,'' in \emph{2006 International Conference on Simulation of Semiconductor Processes and Devices}.\hskip 1em plus 0.5em minus 0.4em\relax IEEE, 2006, pp. 51--54.

\bibitem{ghibaudo2022fluctuation}
G.~Ghibaudo, ``On the fluctuation-dissipation of the oxide trapped charge in a {MOSFET} operated down to deep cryogenic temperatures,'' \emph{arXiv preprint arXiv:2204.04958}, 2022.

\bibitem{enz1996circuit}
C.~C. Enz and G.~C. Temes, ``Circuit techniques for reducing the effects of op-amp imperfections: autozeroing, correlated double sampling, and chopper stabilization,'' \emph{Proceedings of the IEEE}, vol.~84, no.~11, pp. 1584--1614, 1996.

\end{thebibliography}

\end{document}